\documentclass[usegraphicx,useAMS,usenatbib]{mn2e}
\usepackage{graphicx}
\usepackage{psfig,epsfig}
\usepackage{graphics}
\usepackage{epsf}
\usepackage{url}
\usepackage{amssymb}
\usepackage{aas_macros}

\def\Mpc  {{\it h}^{-1}\,{\rm Mpc}}
\def\Mpc{{\it h}^{-1}\,{\rm Mpc}}
\def\Msol {{\it h}^{-1}\,{\rm M_\odot}}
\def\Lsol {\,{\it h}^{-2}\, {\rm L_\odot}}
\def\lcdm{$\Lambda$CDM~}
\def\ie{{\rm i.e. }}
\def\eg{{\rm e.g. }}

\def\d{{\rm d}}
\def\bj{b_{\rm J}}

\def\lsim{\mathrel{\hbox{\rlap{\hbox{\lower4pt\hbox{$\sim$}}}\hbox{$<$}}}}

\title[2dFGRS connected structures]
  {Connected structure in the 2dFGRS}

\author[D.N.A. Murphy, V.R. Eke and Carlos S. Frenk]
  {D.N.A. Murphy\thanks{david.murphy@durham.ac.uk}, V.R. Eke and Carlos S. Frenk\\
Institute for Computational Cosmology, Department of Physics, University of Durham, South Road, Durham, DH1 3LE, UK}

\date{}

\pagerange{\pageref{firstpage}--\pageref{lastpage}} \pubyear{2011}

\begin{document}
\label{firstpage}

\maketitle
\begin{abstract}
We describe and apply a simple prescription for defining connected
structures in galaxy redshift surveys. The method is based upon two
passes with a friends-of-friends groupfinder. The first pass uses a
cylindrical linking volume to find galaxy groups and clusters, in
order to suppress the line-of-sight smearing introduced by the large
random velocities of galaxies within these deep potential wells. The
second pass, performed with a spherical linking volume, identifies the
connected components. This algorithm has been applied to the 2dFGRS,
within which it picks out a total of 7,603 systems containing at least
two galaxies and having a mean redshift less than 0.12. Connected
systems with many members appear filamentary in nature, and the
algorithm recovers two particularly large filaments within the
2dFGRS. For comparison, the algorithm is has also been applied to
$\Lambda$CDM mock galaxy surveys. While the model population of such
systems is broadly similar to that in the 2dFGRS, it does not
generally contain such extremely large structures.
\end{abstract}

\begin{keywords}
cosmology: observations -- large-scale structure of universe
\vspace*{-0.5 truecm}
\end{keywords}

\section{Introduction}
At very large scales, baryonic material is concentrated into an
interconnected sponge-like network known as the Cosmic Web
\citep{1996Natur.380..603B}. Successive redshift surveys have traced
out imposing overdensities in the galaxy distribution. Notable
examples are the CfA `Great Wall' \citep{1989Sci...246..897G}
and the `Sloan Great Wall' \citep{2005ApJ...624..463G}, which was also
noted, if not named, in the Two-degree Field Galaxy Redshift Survey
\citep[2dFGRS,][]{2001MNRAS.328.1039C} by \citet{2004MNRAS.351L..44B}
and \citet{2004MNRAS.352..939E}.

Most studies of large-scale structure in the Universe concentrate on
measuring the galaxy power spectrum \citep[\eg][]{2005MNRAS.362..505C}
or its Fourier transform, the two-point correlation function
\citep[\eg][]{2002ApJ...571..172Z}. These quantities will provide a complete
statistical description of the galaxy distribution provided that their
number density fluctuations are Gaussian. The standard model of
structure formation, $\Lambda$CDM, does assume that the initial
inhomogeneities in the density field, generated during inflation, are
Gaussian. However, the subsequent growth of structure due to
gravitational instability induces significant non-Gaussianities in the
density field, and higher order moments of the density distribution
become important. These phase correlations within the density field
can be characterised either through higher order galaxy correlation
functions 
\citep{2004MNRAS.351L..44B,2004MNRAS.352.1232C,2005MNRAS.364..620G,2006MNRAS.368.1507N}
or through the properties and distribution of filaments.
Quantitative studies of filamentary structures in redshift
surveys have been performed by 
\citet{1983MNRAS.205P..61B,1985MNRAS.216...17B,2003A&A...405..425E,2004MNRAS.347..137P}
and \citet{2010A&A...510A..38S}.

A variety of algorithms have been designed to describe the morphology
of these large-scale structures using different techniques such as
percolation \citep{1983MNRAS.205P..61B}, visual identification of
regions between clusters
\citep{2004MNRAS.354L..61P,2005MNRAS.359..272C}, minimal spanning
trees \citep{1985MNRAS.216...17B,2007MNRAS.375..337C}, the density
field's Hessian
\citep{2007A&A...474..315A,2010MNRAS.406.1609B,2009ApJ...706..747Z},
gradient \citep[Morse theory,][]{2006MNRAS.366.1201N} or linkage
between the two \citep{2008MNRAS.383.1655S,2008ApJ...672L...1S}, the
Hessian of the potential field
\citep{2007MNRAS.375..489H,2009MNRAS.396.1815F}, Delaunay
tessellations \citep{2000A&A...363L..29S,2007A&A...474..315A}, the
Candy and Bisous models
\citep{2005A&A...434..423S,2010A&A...510A..38S} and watershed
transforms \citep{2009MNRAS.393..457S,2010ApJ...723..364A}. Many of
these algorithms also partition the whole of space into clusters,
walls, filaments and voids. They are often applied to dark matter
simulations to help describe the mass distribution, but with a few
notable exceptions
\citep{1983MNRAS.205P..61B,1985MNRAS.216...17B,2004MNRAS.347..137P,2010MNRAS.406.1609B,2010A&A...510A..38S},
they rarely include a comparison with observational data. A primary
motivation for this paper is to carry out a detailed quantitative
comparison of filament properties using the 2dFGRS and mock galaxy
catalogues created using a \lcdm simulation
\citep{2008MNRAS.383..755A} and a semi-analytical galaxy formation
model \citep{2005MNRAS.356.1191B}.

An important aspect of the comparison between model and observed
large-scale structure concerns the existence in both the 2dFGRS and
the Sloan Digital Sky Survey \citep[SDSS;][]{2000AJ....120.1579Y} of
some extremely large structures. These objects are known to have a
large impact on the higher order correlations of the galaxy
distribution
\citep{2004MNRAS.351L..44B,2004MNRAS.352.1232C,2005MNRAS.364..620G,2006MNRAS.368.1507N},
and on its topology \citep{2005ApJ...633...11P}. How common such
structures are within the \lcdm model remains contentious (Yaryura,
Baugh \& Angulo 2010). \nocite{2010arXiv1003.4259Y}

In Section 2, we describe the detection algorithm that we have
developed, and the observational and mock data that will be
compared. The results of this comparison are described in Section 3 
and conclusions drawn in Section 4.

\section{Methods}
\label{method}
In this section, we will describe the observational data that will be
analysed, the algorithm for finding connected systems, the calculation of the
luminosity of such objects, and the mock catalogues that will be used in
order to compare the $\Lambda$CDM model with the observational data.

\subsection{Data}

The observational data used in this study are the two large contiguous
patches towards the north and south galactic poles (NGP and SGP) in
the final data release of the 2dFGRS \citep{2001MNRAS.328.1039C}. In
total, these regions contain 191,897 galaxies with a median redshift
of 0.11, covering an area of approximately 1500 deg$^2$ to a flux
limit corresponding to $b_{\rm J}\sim 19.45$. A bright flux limit of
$b_{\rm J}<14$ is also imposed to exclude objects whose total fluxes
are difficult to determine from APM photographic plates.

\subsection{The detection algorithm}

A desirable property of an algorithm extracting large-scale structure
is that there should be no preferred direction for the resultant
systems. However, redshift space distortions make the line of sight a
special direction in galaxy redshift surveys. The most striking
consequence of non-Hubble flow velocities is to stretch galaxy
clusters, creating 'fingers of god' \citep{1972MNRAS.156P...1J} in the
redshift-space galaxy distribution. These elongated redshift-space
distortions need to be removed before searching for real
structures. We achieve this by taking the 2PIGG catalogue of groups
and clusters constructed by \citet{2004MNRAS.348..866E} from the
2dFGRS. These were found using a friends-of-friends algorithm with a
cylindrical linking volume pointing along the redshift direction as
described in their paper. Having found galaxies belonging to groups
and clusters in this way, we would like to collapse the 'fingers of
god' by placing these galaxies at their group centre positions. One
complication is that \citet{2004MNRAS.348..866E} note that they would
expect the 2PIGG catalogue to contain a few tens of per cent of
interloper galaxies that are incorrectly assigned to groups. To try
and correct for this inevitable misassignment, we choose to retain a
redshift-dependent fraction
\begin{equation}
f(z) = \frac{2-z}{2.4+z}
\label{fz}
\end{equation}
of the members assigned to each group. This expression for $f(z)$ is
derived from the contamination of groups found in mock 2dFGRS
catalogues by \citet{2004MNRAS.348..866E}. The randomly selected
fraction $f(z)$ are all replaced by a single point at the group
centre, whereas the $1-f(z)$ of `interloper' galaxies are jettisoned
from the list of group members and left at their observed redshift
space positions.

The first friends-of-friends pass suppresses the redshift space
distortions associated with intragroup line-of-sight galaxy
velocities. Note that this collapse does not account for the
coherent infall of galaxies onto overdensities that will enhance and merge
structure in the plane of the sky \citep[see
e.g,][]{1987MNRAS.227....1K,1997ApJ...479L..15P}. We then apply a
friends-of-friends algorithm with a spherical linking volume to the
set of remaining galaxies and group centres. The radius of this
linking sphere is chosen to be $b$ times the mean intergalactic
separation at that redshift, as defined in Eqn.~2.7 of
\citet{2004MNRAS.348..866E}. Small linking lengths would lead to many
small systems, whereas very large ones would lead to percolation, and
a single large connected component encompassing everything in the
survey. An intermediate value for $b$ will lead to a more useful
description of the structures present in the survey. This two-pass
procedure provides a new and simple way to define connected structures
in galaxy redshift surveys.

\begin{figure}
\centering
\centerline{\psfig{file=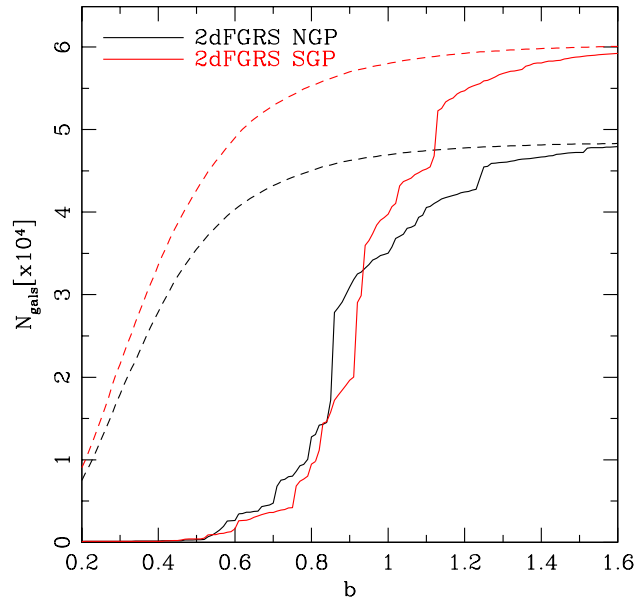,height=8.5cm}}
\caption{Variation in the number of 2dFGRS galaxies in systems
  extracted by the algorithm as the relative linking length, $b$,
  changes. Black lines denote galaxies from the NGP, red lines from
  the SGP. Solid lines represent the number of galaxies in the largest
  structure, whilst the dashed lines show the number of galaxies in
  all systems with at least two members.
\label{perc}}
\end{figure}

\begin{figure}
\centering
\centerline{\psfig{file=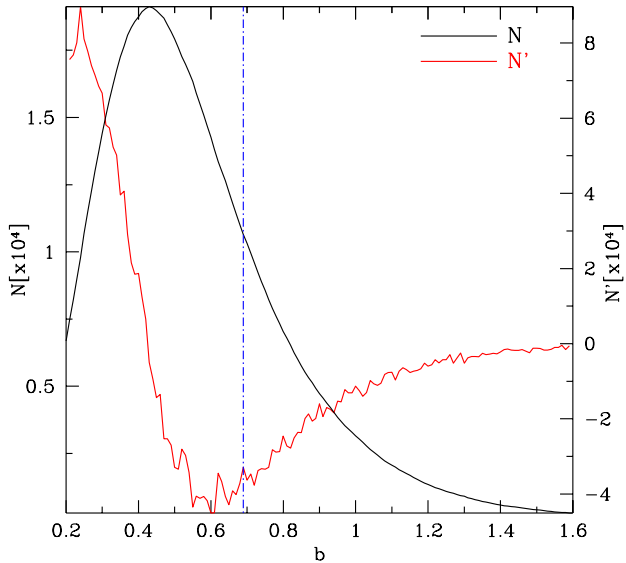,height=8.5cm}}
\caption{Variation in the number of structures extracted by the
algorithm as the relative linking length, $b$, varies. The red line
represents the first derivative of this function, corresponding to the
rate of change of system number. We adopt a relative linking
length, $b=0.69$, close to this minimum, as denoted by the
dot-dashed blue line. This corresponds to systems approximately
bounded by a surface with a galaxy number overdensity of $\sim 1.5$.
\label{nfil}}
\end{figure}

Fig.~\ref{perc} shows how the number of galaxies in connected
structures and the number of galaxies in the largest system vary with
$b$ for both the NGP and SGP. Both the NGP and SGP regions show a
rapid growth of their largest system as $b$ increases beyond $\sim
0.8$. Fig.~\ref{nfil} shows the variation of the total number of
connected structures and its first derivative. We pick
$b=0.69$ as a value that gives rise to an interestingly large range of
system sizes. This corresponds to finding structures bounded by an
irregular surface that has an overdensity of
$\Delta\rho/\bar{\rho}=3/(2\pi b^{3})\approx 1.5$ \citep{1996MNRAS.281..716C}.

This choice is close to the point at which $\d N/\d b=0$, where the growth in the number of systems
arising from single galaxies becoming linked matches the decrease
caused by merging the structures together. The resulting systems are
shown in Fig.~\ref{twodfrich}. 

The abundance and extent of survey-sized connected structures will depend upon
the geometry of the survey to which this algorithm is applied. Thus,
while this technique is appropriate for comparing an observed data set
with a mock catalogue of that particular survey, care is required when
trying to infer the physical properties and abundance of the largest
structures in the underlying distribution.

\subsection{Connected structure luminosities}

\begin{figure*}
\centering
\includegraphics[width=\linewidth]{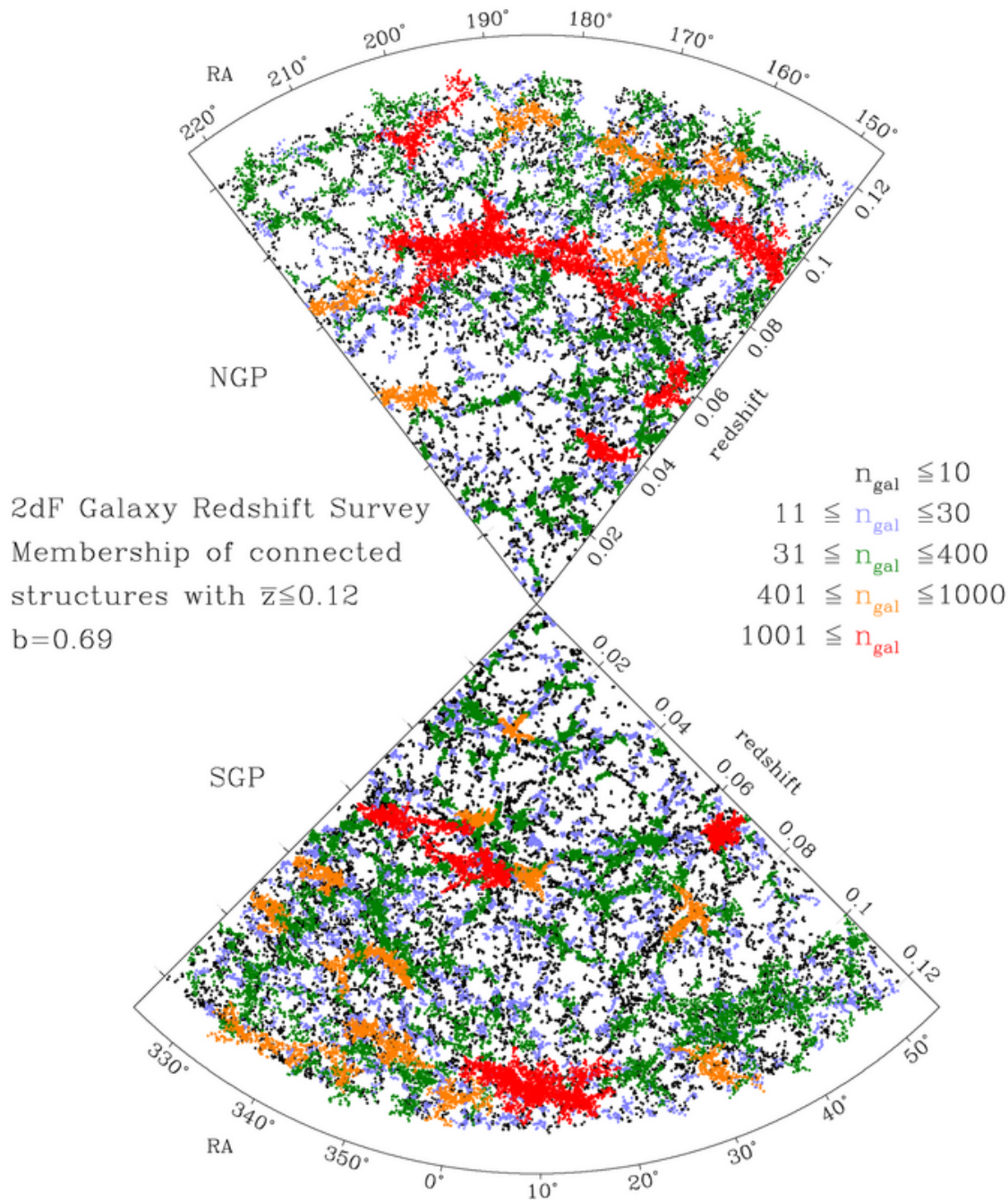}
\caption{Spatial distribution of 2dFGRS galaxies in connected
  structures for systems with average redshift $\bar{z}\leq 0.12$ in
  the RA-$z$ plane. These systems contain at least two galaxies and
  dot colours represent the weighted number of galaxies in the
  structure, where this weight takes into account the local angular
  incompleteness.
\label{twodfrich}}
\end{figure*}

\begin{figure*}
\centering
\includegraphics[width=\linewidth]{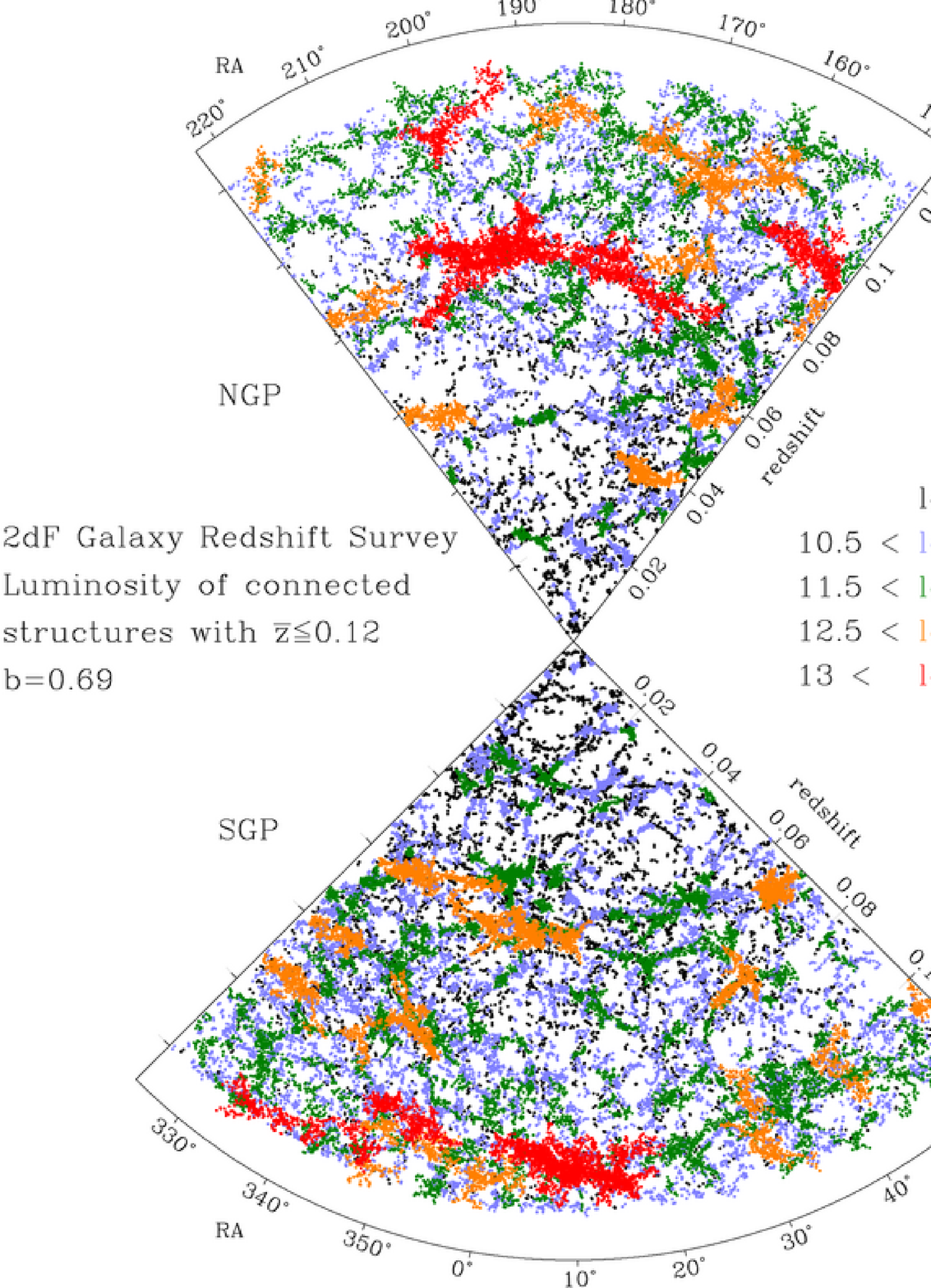}
\caption{Spatial distribution of 2dFGRS galaxies in connected
  structures in the RA-$z$ plane. Colours represent the total system
  luminosity in units of ${\rm log}_{10}(L/h^{-2}L_{\odot})$.
\label{twodflum}}
\end{figure*}

We would like to quantify the sizes of the objects found using this
method in a way that (a) does not depend explicitly on the magnitude
limit of the survey and (b) assigns the same size to a particular
structure, independently of the redshift at which it is placed. Thus,
rather than merely counting the number of galaxies present in each
system, we define a luminosity that takes into account the flux limits
of the survey. The angular variation of the flux limit in the 2dFGRS
is such that it changes over the length of the elongated filamentary
structures. Consequently, it is necessary to convert the observed
luminosity of each galaxy to the total luminosity that would have been
seen without any flux limits, rather than correcting the observed
luminosity of the system as a whole. This is done assuming that the
galaxy luminosity function $\Phi(L)$ is given by the Schechter
function determined by \citet{2002MNRAS.336..907N},
i.e. $(L_*,\alpha)=(10^{10}\Lsol, -1.21)$ and using
\begin{equation}
\frac{L_{\rm cor}}{L_{\rm obs}}=\frac{\int_{0}^{\infty}L \Phi(L) dL}{\int_{L_{min}}^{L_{max}}L \Phi(L) dL},
\end{equation}
where the luminosity limits in the integral in the denominator reflect
the upper and lower flux limits evaluated at the redshift of the
galaxy. $L_{\rm cor}$ and $L_{\rm obs}$ represent the corrected and
observed galaxy luminosities respectively, taking into account the
$k+e$ correction in a manner similar to \citet{2002MNRAS.336..907N}:
\begin{equation}
k+e=\frac{z+6z^2}{1+8.9z^{2.5}}.
\end{equation}
The total luminosity is calculated by summing up all the
corrected galaxy luminosities for galaxies within that system, 
taking into account the weighting factors that describe the local
incompleteness of the survey. Given
the flux limit of the 2dFGRS, the fraction of the total luminosity
that is observed drops beneath a half at redshifts exceeding $z \sim
0.12$. For this reason, we will restrict our analysis to structures
with $\bar{z}\le0.12$. Fig.~\ref{twodflum} shows the systems found
within the 2dFGRS colour coded according to their
luminosity. Comparing with Fig.~\ref{twodfrich}, it is apparent how
the luminosity picks out structures at larger distances than the
membership, which includes only galaxies within the flux limits.

\subsection{Mock catalogues}

In order to address how well the observed distribution of system
luminosities compares with that predicted for a 2dFGRS-like survey of
a \lcdm cosmological model, we need to create mock galaxy
surveys. This is done using a combination of the BASICC dark matter
simulation described by \citet{2008MNRAS.383..755A}, and the
semi-analytical model of \citet{2005MNRAS.356.1191B}, which is a
development of that introduced by \citet{2000MNRAS.319..168C}.

In brief, the BASICC simulation contains $1448^3$ particles in a
$1340\Mpc$-long cube of a \lcdm model with $\Omega_{\rm M}=0.25$,
$\Omega_{\Lambda}=0.75$, $\Omega_{\rm b}=0.045$, $h=0.73$ and
$\sigma_8=0.9$. \citet{2008MNRAS.383..755A} used a friends-of-friends
(FOF) algorithm with a linking length of $0.2$ times the mean
interparticle separation to define haloes. Requiring 10 particles to
resolve a halo implies that the minimum resolvable halo mass is
$5.5\times 10^{11}\Msol$. All of these haloes were populated with
galaxies according to the semi-analytical model of
\citet{2005MNRAS.356.1191B}. Galaxies that reside in unresolved haloes
are randomly placed onto dark matter particles outside resolved
haloes, according to the method described by
\citet{2005MNRAS.362..505C}.

While the galaxy luminosity function produced by this model is close
(within 0.3 magnitudes around $L_*$) to that observed in the 2dFGRS
\cite{2002MNRAS.336..907N}, we nevertheless apply a
luminosity-dependent shift in luminosities so that the cube of model
galaxies has exactly the same luminosity function as the 2dFGRS. Only
galaxies with $L>1.4\times 10^8\Lsol$ are included in the model
cube. For the flux limit of the 2dFGRS, this implies that the mock
galaxy catalogues will be missing low luminosity galaxies at $z\lsim
0.025$. This corresponds to just under one per cent of the total
volume being considered.

Fifty mocks of the 2dFGRS were created from this cube of model
galaxies as follows:\\
a) a random observer location and direction were chosen,\\
b) volume-limited NGP and SGP surveys were created using periodic
replicas of the cube if required. Note that the depth of the 2dFGRS is
less than half the length of the BASICC simulation cube, and its
effective volume is less than $10^{-3}$ of the volume of
the BASICC simulation,\\
c) galaxies were removed according to the position-dependent 2dFGRS
flux limits and completeness masks,\\
d) the remaining galaxies were assigned a redshift according to 
their peculiar velocity in the simulation.\\

Previous studies have shown that this semi-analytic model tends to
place too many low luminosity galaxies into galaxy clusters
\citep{2004MNRAS.355..769E,2008MNRAS.385L.116G,2009MNRAS.400.1527K}.
Given that the global luminosity function has been forced to match
that in the observations, this implies that the model will lack low
luminosity galaxies in lower density regions. This known problem will
affect the structure finder.

To try and reduce the impact of this known difference between the
model and observations, we allow ourselves the freedom to jettison a
smaller fraction of galaxies from the model groups than given by
Eqn.~\ref{fz} for the real 2dFGRS. This decreases, in the vicinity
of the groups, the number density of points used for the
structure-finding sweep of the friends-of-friends algorithm to an
amount similar to that in the real survey. We achieve this in the
model by multiplying $f(z)$, as given by Eqn.~\ref{fz}, by a
constant, $\chi$. $\chi>1$ implies that a higher fraction of galaxies
are retained in the groups. 

In order to determine an appropriate value for $\chi$, we have measured
the distribution of system orientations defined as
\begin{equation}
\theta={\rm tan}^{-1}\left(\frac{\Delta l_z}{\Delta l_\phi}\right),
\end{equation}
where $\Delta l_z$ represents the range of the member galaxies in the
redshift direction and $\Delta l_\phi$ is the larger of the ranges of
member galaxies in the RA and dec directions. Thus, $\theta=\pi/2$ for
a radial object and $0$ for one lying perpendicular to this. We use
the greater of $\Delta l_z$ and $\Delta l_\phi$ to describe the scale
size of the connected structure.

\begin{figure}
\centering
\centerline{\psfig{file=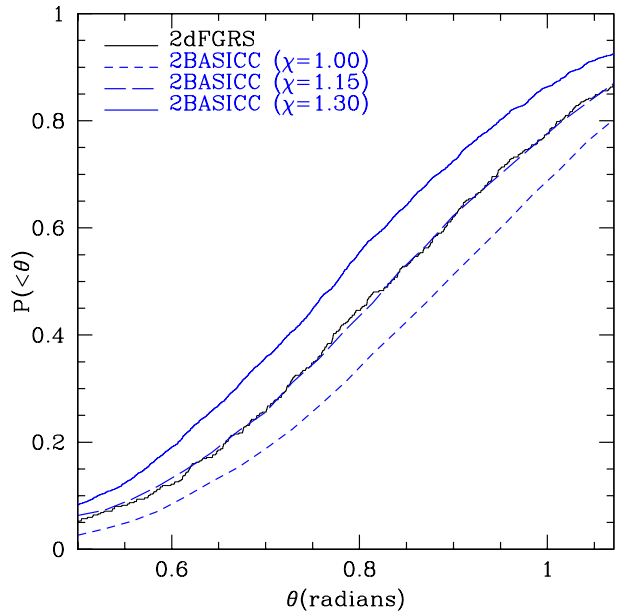,height=8.5cm}}
\caption{Cumulative probability distributions of system orientations
  for all objects containing at least 20 galaxies. Results are shown
  for the 2dFGRS and for averages of 10 2BASICC mocks. The mock
  distributions are derived from three different choices of
  $\chi=1.0,1.15$ and $1.30$, as indicated in the legend.
\label{chiang}}
\end{figure}

Fig.~\ref{chiang} shows the cumulative probability distributions of
system orientations for structures containing at least 20 galaxies in
the 2dFGRS and those recovered from 10 mock surveys using three
different values of $\chi$\footnote{One might imagine that randomly
  oriented connected structures would be uniformly distributed with
  $\theta$. However, since systems often contain more than two
  galaxies, which are generally not colinear, the definition of
  $\theta$ leads to connected structures preferentially avoiding
  values towards the ends of the range $[0,\pi/2]$.}. It is apparent
that, when treated in the same way as the real data (\ie $\chi=1$),
the model contains too many objects aligned along the line of
sight. This is a result of too many low luminosity galaxies being
placed into the redshift space volumes occupied by the model
groups. When the `interloping' galaxies are jettisoned from the groups
found in the mock catalogues, enough of these extra low luminosity
galaxies are placed along radial lines that they bias the orientation
distribution. Increasing $\chi$ retains a higher fraction of the
initially grouped galaxies in the groups, reducing the number of
`interlopers' returned to the field, and decreasing the number of
radially aligned objects found in the second pass of the
friends-of-friends algorithm. A value of $\chi=1.15$ produces a mock
orientation distribution that is, according to a K-S test,
indistinguishable from that found in the 2dFGRS. This is chosen as the
default value for these 2BASICC mocks throughout this paper.

An additional set of 22 mock 2dFGRS surveys created from the Hubble
Volume by \citet{2000MNRAS.319..168C}, as used by
\citet{2005MNRAS.364..620G}, are also analysed to give some idea of
the systematic differences resulting from different simulations and
implementations of the galaxy formation modelling. For these mock
catalogues, a value of $\chi=1.11$ was required in order to recover
the 2dFGRS system orientation distribution.

\begin{table*}
\centering \caption[] { Properties of connected structures with $\bar{z}\leq0.12$
 identified in the 2dFGRS and mock surveys. Mock values are the mean
 over all 50 2BASICC and 22 Hubble Volume mock surveys with the
 uncertainties being the standard deviation of individual surveys from
 these mean values. $N$ is the total number of connected systems
 within the catalogue, $f$ the fraction of galaxies out to
 $z=0.12$ in systems and $N_{\rm gal}$ is the total number of
 galaxies out to $z=0.12$. The fifth and sixth columns describe the
 average and maximum object luminosities (in units ${\rm
 log}_{10}(L/h^{-2}L_{\odot})$). We give the comoving scale $l_{max}$
 of the largest structure identified in the survey in the final
 column. This is defined as the largest in extent of galaxy members in
 the redshift, RA or dec directions.
\label{stats}}
\begin{center}
\begin{tabular}{|l|c|c|c|c|c|c|}
Survey & $N$ & $f$ & $\log_{10}(N_{\rm gal})$ & $\log_{10}\bar{L}$ & $\log_{10}L_{max}$ & $l_{max} (\Mpc)$ \\
\hline
2dFGRS & 7603 & 87.7\% & 5.06 & 11.16 & 13.89 & 198\\
2BASICC & $8023\pm250$ & $85.9\pm0.8\%$ & $5.07\pm0.04$ & $11.08\pm0.03$ & $13.44^{+0.15}_{-0.23}$ & $81\pm19$\\
HV & $8253\pm135$ & $82.0\pm0.8\%$ & $5.11\pm0.02$ & $11.10\pm0.03$ & $13.55^{+0.14}_{-0.20}$ & $93\pm27$\\
\end{tabular}
\end{center}
\end{table*}

\section{Results}
\label{results}

In this section, we will first describe the main features and
properties of the connected structures found in the 2dFGRS and then compare 
with the results from the \lcdm mock surveys. This comparison will
encompass both the whole population of connected systems as well as the
largest objects.

\subsection{Connected systems in the 2dFGRS}

\begin{figure}
\centering
\centerline{\psfig{file=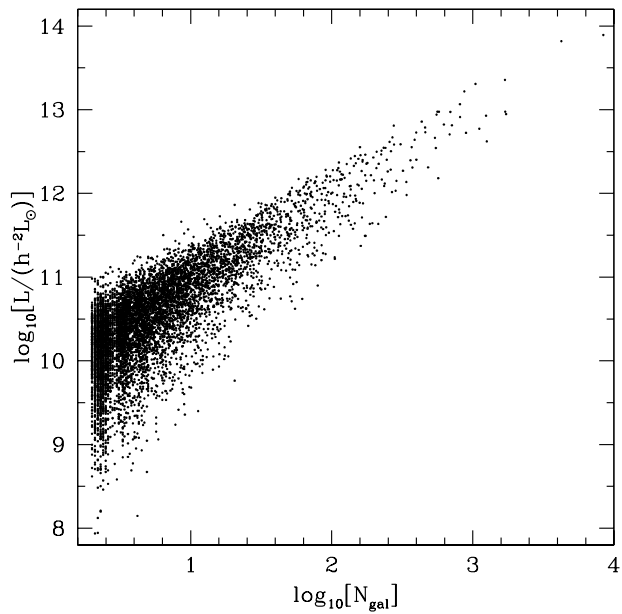,height=8.5cm}}
\caption{The relation between object luminosity, $L$, and
$N_{gal}$, the weighted number of galaxies in objects with
$\bar{z}\leq 0.12$.
\label{lumrich}}
\end{figure}

Projections onto the right ascension-redshift plane of all of the
connected systems found in the 2dFGRS are shown in
Figs.\ref{twodfrich} and \ref{twodflum}. A total of $95,010$ galaxies
are linked into $7,603$ systems containing at least two members and
mean redshifts no greater than $0.12$. Of these, $3,018$ contain only
two members. Almost $87$ per cent of galaxies at $z\le0.12$ are
placed into a connected structure. One large filamentary-structure
stands out in each of the NGP and SGP wedges. These systems trace out
the same overdensities apparent in the 2PIGG distribution
\citep{2004MNRAS.355..769E}, the smoothed galaxy density map
\citep{2004MNRAS.351L..44B} and the reconstructed density field
\citep{2004MNRAS.352..939E} of the 2dFGRS. The largest NGP object, at
$z\sim0.08$, corresponds to the large RA end of the `Sloan Great Wall'
highlighted by \citet{2005ApJ...624..463G}. At a total $\bj$-band
luminosity of $\sim 7.8\times10^{13}\Lsol$, this is about $20$ per
cent more luminous than the largest equivalent in the SGP, which lies at
$z\sim0.11$ and RA $\sim10^\circ$. The extents in RA of these largest
NGP and SGP systems in comoving coordinates are $\sim198\Mpc$ and
$99\Mpc$ respectively. While the NGP systems contains twice as many
members as that in the SGP, it is very nearly broken into two pieces
around RA $\sim185^\circ$, where the galaxy density drops off
considerably.

More locally, a continuation to lower declinations of the CfA Great
Wall \citep{1989Sci...246..897G} is seen at $z\sim0.02$ in the NGP,
although our algorithm breaks this up into a few different components.

\begin{figure}
\centering
\centerline{\psfig{file=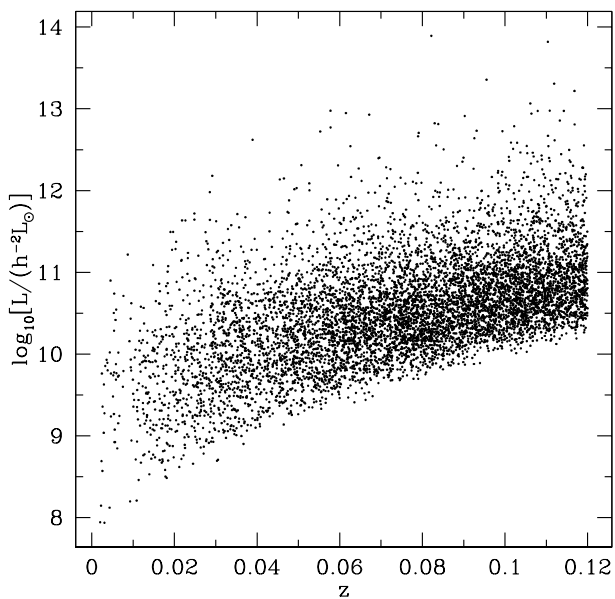,height=8.5cm}}
\caption{The distribution of system luminosities with increasing
  redshift.
\label{lumzdist}}
\end{figure}

Some average and extreme properties of the systems identified in the
2dFGRS are listed in Table~\ref{stats}. In more detail, the
correlation between the luminosity and weighted (to account for the
local angular incompleteness of the survey) membership of connected
structures is shown in Fig.~\ref{lumrich}. The second largest system
contains at least twice as many members as the third largest one, and
almost 3 times as much luminosity, making the largest NGP and SGP
structures stand out from the remaining systems. The scatter around
the mean relation reflects the range of redshifts in the flux-limited
survey. While the object luminosity is corrected to take account of
this flux limit, the weighted number of galaxies is not.

Fig.~\ref{lumzdist} shows how connected system luminosity varies with
redshift, with the lower envelope representing the total luminosity of
2 galaxies at the flux limit. The geometry of the survey precludes
finding very luminous structures at low redshift because of the small
volume sampled, but the greater volume available at
larger redshifts is sufficient to contain larger, more luminous
filamentary structures. The largest NGP and SGP systems are once again 
conspicuous at the top of the figure.

\subsection{Comparison with mock catalogues}

\begin{figure}
\centering
\centerline{\psfig{file=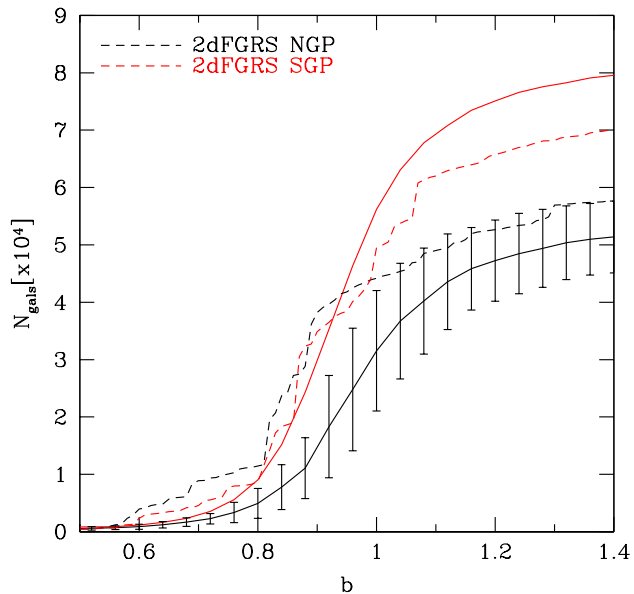,height=8.5cm}}
\caption{The weighted number of galaxies in the largest structure for
  NGP (black) and SGP (red) systems with $\bar{z}\leq 0.12$,
  including any members with redshifts greater than this
  limit subject to the mean system redshift remaining below it. In
  both cases the dashed lines show the 2dFGRS data, whilst solid lines
  represent the mean number of galaxies in the largest object across
  50 mock surveys. In the NGP case, we include also error bars
  representing the standard deviation of these surveys around the
  mean.
\label{2bperc}}
\end{figure}

\begin{figure}
\centering
\centerline{\psfig{file=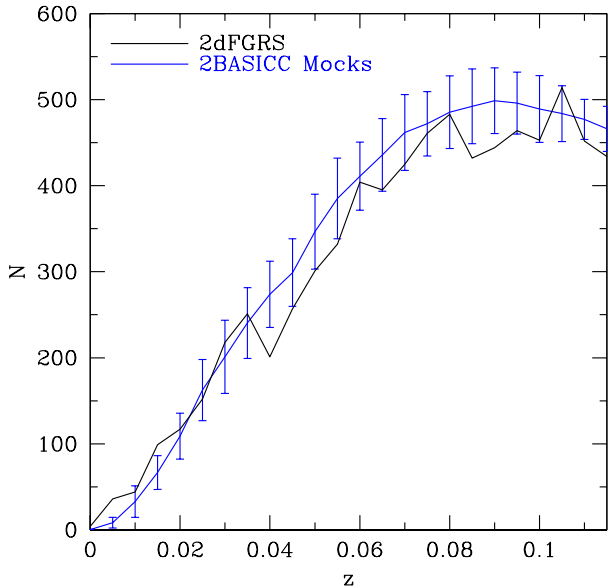,height=8.5cm}}
\caption{The redshift distribution of all objects with at least two
  members. The blue line represents the mean number of connected
  structures as a function of redshift across 50 mock surveys, with
  error bars representing their standard deviation around the
  mean. The black line corresponds to systems detected in the 2dFGRS.
\label{2bnumzdist}}
\end{figure}

We follow almost the same procedure to define connected structures in the 50
2BASICC mocks, with the only difference being that the fraction of
galaxies retained in the groups, $f(z)$ in
Eqn.~\ref{fz}, is increased by a factor $\chi=1.15$, as described
in \S\ref{method}. The impact of this choice on the results is discussed
later.

Fig.~\ref{2bperc} shows how the number of galaxies in the most
populated system grows as $b$ is increased for each survey
region. While the behaviour is broadly similar to that of the largest
filament in the 2dFGRS, the onset of percolation in mock catalogues is
delayed by about $0.1$ times the mean intergalaxy separation. As a
consequence, at $b=0.69$, the largest 2dFGRS system
(located in the NGP) is a significant outlier, being more populated
than the corresponding structure in all but one of the 50 mock
catalogues. The mock system containing more galaxies than the
largest one in the 2dFGRS is placed at $z \sim 0.01$ and is much less
luminous. It is the result of the randomly chosen observer being put
very near to a large galaxy cluster.

The larger value of $\chi$ adopted for mock surveys means that the
number of galaxies and group centres used for the algorithm is
on average $\sim 13$ per cent lower than in the case of the 2dFGRS. If
we were to use a value of $b$ that was correspondingly
larger (\ie scaled by the inverse cube root of the number of points to
be $0.72$) then this does not significantly affect the discrepancy
between the number of galaxies in the most populated systems in the
mock or real 2dFGRS.

The redshift distribution of the structures is shown in
Fig.~\ref{2bnumzdist}. At $z\lsim0.025$, where the mock catalogues are
known to be missing low luminosity galaxies, the mock surveys contain
fewer objects than the real 2dFGRS. The main reason for this is
actually not the incompleteness in the mocks, but the fact that too
many low luminosity galaxies are placed into large groups, reducing
the number available to form other small systems. This local volume
represents only a small fraction of the survey.

For redshifts greater than $0.04$, the number of real 2dFGRS
structures is typically slightly below the mean of the 50 mock
surveys. This is reflected in the first column of Table~\ref{stats},
which shows that the total number of systems in the 2dFGRS is
$(1-2)\sigma$ beneath that of the mocks, despite the fact that a
slightly higher fraction of galaxies are placed into the 2dFGRS structures.
The total number of galaxies in the 2dFGRS and mock surveys matches
well by construction, but the excess of mock galaxies placed into
groups means that fewer are available in lower density regions for
linking together small systems.
The relatively high fraction of galaxies placed into structures and low
number of structures in the 2dFGRS leads to a larger mean 
luminosity. This difference can be removed by not including the two most
luminous systems in the 2dFGRS in the calculation.

\begin{figure}
\centering
\centerline{\psfig{file=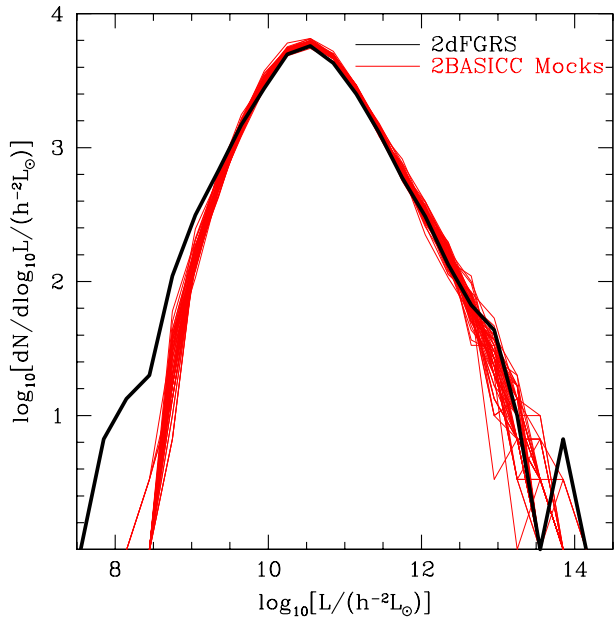,height=8.5cm}}
\caption{The distribution of luminosities for all structures with a
minimum membership of two, out to a redshift $z=0.12$. The red lines
show the distribution for each of the 50 2BASICC mock catalogues. The
black line shows the distribution for systems in the 2dFGRS.
\label{2blumdist}}
\end{figure}

The distribution of system luminosities is shown in
Fig.~\ref{2blumdist}. The mocks have a relative lack of structures
beneath $L\sim10^9\Lsol$, more than the real 2dFGRS at
$L\sim3\times10^{10}\Lsol$, which is the peak of the
distribution and corresponds approximately to two $L_*$ galaxies, and
a paucity of filamentary systems like the largest ones in the
2dFGRS. As stated above, the difference between the model and real
distributions at low luminosities arises mostly because the lowest
luminosity galaxies in the model are more likely to be placed into
larger groups and hence are not available to form very low luminosity
systems. The deficit of lower luminosity galaxies outside groups
impacts in two ways upon the most luminous model structures. They tend
to gain luminosity because their groups are slightly more luminous
than those of corresponding mass in the 2dFGRS. However, the lack of
low luminosity galaxies in the lower density regions, makes it less
likely that large structures will join together. It is this second
effect that is more important, resulting in none of the 50 mock
surveys yielding two systems as luminous as the two most luminous in
the 2dFGRS.

\begin{figure}
\centering
\centerline{\psfig{file=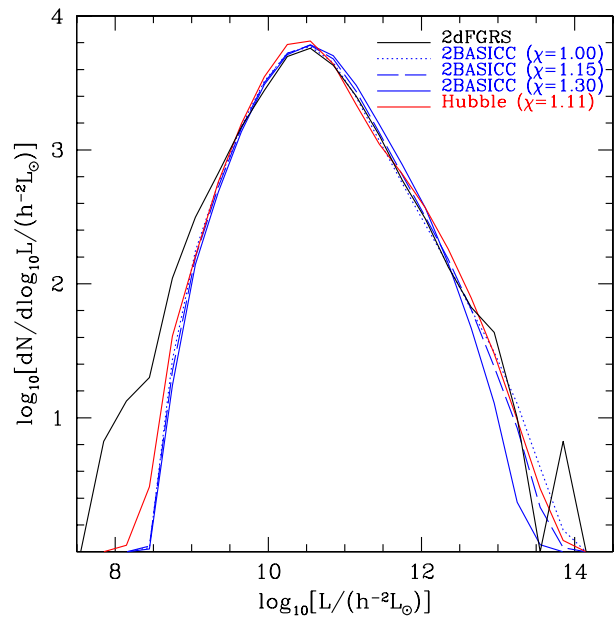,height=8.5cm}}
\caption{Structure luminosity distributions for different values of
  $\chi$ in the 2BASICC surveys, for the Hubble Volume catalogues and
  the 2dFGRS. Mock survey distributions have been averaged over the 50
  surveys in the 2BASICC and the 22 in the Hubble Volume simulations.
\label{dist_hv_chi}}
\end{figure}

Given that many differences between the real and mock structure
luminosity distributions result from the different spatial
distributions of low luminosity galaxies in the real and mock surveys,
and that we have used a different value of $\chi$ for real and mock
surveys, one might reasonably ask what changes when $\chi=1$ is used
for the mocks. This is shown in Fig.~\ref{dist_hv_chi}, where three
different $\chi$ values are used for the 2BASICC mocks. Increasing
$\chi$ retains more galaxies in the groups, leaving fewer galaxies to
help the algorithm link together larger structures. This leads to a
decrease in the luminosity of the most luminous systems. Decreasing
$\chi$ leads to an increase in the luminosity of the most luminous
systems, but even for $\chi=1$ there are still no surveys with two
structures at least as luminous as the second most luminous 2dFGRS
system. Nevertheless, we do obtain two surveys with a connected system
more luminous than the brightest 2dFGRS system. However, as shown in
Fig.~\ref{chiang}, this comes at the expense of producing a set of
objects that are significantly more radially oriented than those found
in the 2dFGRS.

Also shown in Fig.~\ref{dist_hv_chi} is the distribution of systems
found in the 22 Hubble Volume mock catalogues of
\citet{2000MNRAS.319..168C}. After tuning $\chi$ to be $1.11$ to
recover the 2dFGRS structure orientation distribution, the Hubble
Volume mocks are broadly similar to the 2BASICC ones, with the
abundance of the most luminous systems being almost unchanged. The
bottom row of table~\ref{stats} contains statistics for the systems
found in these Hubble Volume mocks.

\section{Conclusions}
\label{conclusion}
We have described a simple algorithm with which to define connected
structure within galaxy redshift surveys, and applied it to the
2dFGRS. This algorithm explicitly addresses the redshift-space
distortion associated with rapidly moving galaxies within groups and
clusters. The $7,603$ 2dFGRS connected structures at $z\leq0.12$
containing at least two members range up to $\sim 200\Mpc$ in extent,
but are mostly associations of two $L_*$ galaxies. Quantifying object
sizes via their total luminosities, corrected for the survey flux
limits, we find that the largest systems are filamentary in nature and
have $\bj$ luminosities of almost $10^{14}\Lsol$.

Applying the same algorithm to \lcdm mock 2dFGRS catalogues,
constructed using large-volume dark matter simulations and the
semi-analytical model of \citet{2005MNRAS.356.1191B}, we find a
broadly similar distribution of structures to those in the real
data. There are, however, a few differences in detail. Many of these
result from the fact that the model places too many $L\lsim L_*$
galaxies into groups and clusters compared with the 2dFGRS. This
biases the orientation distribution of the systems containing at least
$20$ galaxies to contain more radially aligned objects in the mock
survey than in the 2dFGRS. Applying a crude correction to the
algorithm to enable it to recover the same orientation distribution in
the mock survey as it does in the 2dFGRS leads to the largest mock
structures being significantly less luminous than those in the 2dFGRS.

It is clear that at least some of the differences between the
properties of the structures in the 2dFGRS and the mock catalogues
arise from inadequacies in the galaxy formation model that was used to
construct the mocks. We have attempted to overcome these inadequacies
as far as possible through empirical corrections. Our analysis
indicates that the largest filamentary structures seen in the 2dFGRS
are not reproduced in the mock catalogues. However, while this
discrepancy could signal a failure of the standard \lcdm cosmological
model on large scales, it seems more plausible that it reflects a
shortcoming in the predictions of our models of galaxy formation for
the abundance and spatial distribution of galaxies on small scales.

\section*{Acknowledgements}
DNAM acknowledges an STFC PhD studentship and CSF a Royal Society
Wolfson Research Merit award. We thank Raul Angulo and Carlton Baugh
for assistance in the production of the 2BASICC mock surveys. We also
thank Kevin Pimblett and the referee for useful comments that
improved the quality of this manuscript.

\bibliography{./filaments.bbl}

\label{lastpage}

\newpage
\end{document}